\documentclass[10pt,twocolumn,table]{article} 
\usepackage{simpleConference}
\usepackage{times}
\usepackage{graphicx}
\usepackage{amssymb}
\usepackage{url}
\usepackage[colorlinks=true,breaklinks=true]{hyperref} 
\usepackage{pgfplots}
\usepackage{pgfplotstable,booktabs}
\pgfplotsset{compat=1.14}
\usepackage{subfig}
\usepackage{graphicx}
\usepackage{todonotes}
\usepackage{siunitx}
\usepackage{enumitem}
\pgfplotstableread[col sep=comma]{data.csv}{\mydata}

\newcommand{\data}[2]{\pgfplotstablegetelem{#2}{#1}\of\mydata\pgfplotsretval ~}

\begin{document}

\title{Long term availability of raw experimental data in experimental fracture mechanics}

\author{Patrick Diehl$^{1,*}$, Ilyass Tabiai$^1$, Felix W. Baumann$^{2,3}$, Martin Levesque$^1$ \\
\\
1. Department for Multi scale Mechanics, Polytechnique Montreal \\
2. TWT GmbH Science \& Innovation\\
3. Institute of Computer-aided Product Development Systems, University of Stuttgart \\
* patrick.diehl@polymtl.ca (https://orcid.org/0000-0003-3922-8419)\\
\today
}

\maketitle
\thispagestyle{empty}

\begin{abstract}
Experimental data availability is a cornerstone for reproducibility in experimental fracture mechanics, which is crucial to the scientific method. This short communication focuses on the accessibility and long term availability of raw experimental data. The corresponding authors of the eleven most cited papers, related to experimental fracture mechanics, for every year from $2000$ up to $2016$, were kindly asked about the availability of the raw experimental data associated with each publication. For the $187$ e-mails sent: $22.46$\% resulted in outdated contact information, $57.75$\% of the authors did received our request and did not reply, and $19.79$ replied to our request. The availability of data is generally low with only $11$ available data sets ($5.9$\%). The authors identified two main issues for the lacking availability of raw experimental data. First, the ability to retrieve data is strongly attached to the the possibility to contact the corresponding author. This study suggests that institutional e-mail addresses are insufficient means for obtaining experimental data sets. Second, lack of experimental data is also due that submission and publication does not require to make the raw experimental data available. The following solutions are proposed: ($1$) Requirement of unique identifiers, like ORCID or ResearcherID, to detach the author(s) from their institutional e-mail address, ($2$) Provide DOIs, like Zenodo or Dataverse, to make raw experimental data citable, and ($3$) grant providing organizations should ensure that experimental data by public funded projects is available to the public.  
\end{abstract}

\section{Introduction}
\label{sec:introduction}
Reproducibility is the ability to obtain the same research results as another researcher, given the same analysis is done on the same raw data. Reproducibility is crucial to the scientific method~\cite{noauthor_scientific_2017,peng2011reproducible}. Reproducibility in experimental mechanics, can hardly be achieved without access to the raw data used by fellow researchers in their publications. The lacking of scientific reproducibility has been shown for basic and preclinical research~\cite{begley2015reproducibility} and psychological science~\cite{open2015estimating}. In biology~\cite{vines2014availability}, a study revealed that raw data sets could be obtained from $19$\% of $516$ papers containing experimental data and published from $1991$ to $2011$.\\

\noindent
Different stakeholders addressed the lacking availability of experimental data. The organization for economic co-operation and development (OECD) was commissioned by different governments to develop a set of guidelines to provide cost-effective access to publicly funded research data~\cite{OECD2007,OECD2004}. Publishers are currently investigating means to strengthen data-access practices~\cite{NATUREEditorial2014} or support open data~\cite{Finneg2015}.\\

\noindent
According to a recent study, more than $70$\% of researchers, out of more than $1500$ polled, have tried and failed to reproduce another scientist's experiments~\cite{baker20161}. However, the study also shows that physicists and engineers are confident that peer reviewed published data is reproducible. \\

\noindent
Most publications in experimental fracture mechanics rely on data gathered from experiments. It has been our experience (and practice) that only quantities of interest are presented and the raw experimental data is usually missing. Moreover, even when experimental data is available, information related to the experimental setup itself is usually sparse (e.g.\,, calibration of the measurement unit, software used, etc.), which prevents the experiment's replication.\\

\noindent
The modelling community is also highly interested in high fidelity and well documented experimental data to validate model predictions~\cite{zhuang_recent_2012}. The data published in the literature usually lacks information about boundary conditions, etc.\,, to ensure that the models reproduce, at least conceptually, the experiments they aim to reproduce.\\

\noindent
This short communication focuses on the accessibility and long term availability of raw experimental data, as well as supporting information, in experimental fracture mechanics. We have contacted the authors of the eleven most cited papers related to experimental fracture mechanics for every year from $2000$ up to $2016$. We kindly asked these authors about the availability of raw experimental data associated with each publication. The up-to-dateness of the e-mail addresses were studied and the reply-behaviors for working e-mail addresses were considered. Finally, the availability of raw data out of the positive responses from the authors was emphasized.\\

\noindent
Section~\ref{sec:results} deals with the methodology for this study and the data collection. Section~\ref{subsec:received_data} presents the main results. In Section~\ref{sec:discussion}, discusses the results and Section~\ref{sec:conclusion_and_outlook} concludes the paper.

\section{Methodology}
\label{sec:results}

\subsection{Data collection}
\label{subsec:collection_of_data}
The \emph{Web of Science} database\footnote{\url{https://webofknowledge.com}} was queried with the following fields on September $18$\textsuperscript{th} $2017$:\\
\texttt{TOPIC: (CRACK) AND TOPIC: (DAMAGE) AND TOPIC: (EXPERIMENTAL) AND YEAR PUBLISHED: ($X$)}, where $X$ varied from $2000$ to $2016$.\\

\noindent
The top eleven cited papers containing experimental data generated by the authors, included as a reference to an online resource or as an appendix, were selected for each of the respective publication year.\\

\noindent
We investigated: \textit{(1)} the document's meta-data provided by \emph{Web of Science} (WoS), \textit{(2)} the PDF document itself and finally \textit{(3)} the publisher's website to identify the corresponding authors email contacts. Less than $25$\% of the papers published between $2000$ and $2004$ contained an e-mail address in the meta data from WoS while $80$\% of the papers published after $2004$ contained that information. The first corresponding author was selected for the communication attempt. The full list of references investigated can be found on Github under the BibTEX format\footnote{\url{https://github.com/OpenDataExpMechanics/Survey}}.\\ 

\noindent
The generic email detailed in~\ref{appendix:letter1} was sent to the selected authors on October 16\textsuperscript{th} 2017. We asked the authors if they were willing to share their experimental data and if so, how long, in minutes, would it take to gather. \\
The prepared prescribed answers were: \textit{(a)} data is not available, \textit{(b)} the data is confidential, \textit{(c)} one of the co-authors should be contacted to obtain the data. Furthermore, an open answer was available to the authors that were unable, or unwilling, to share this information.\\

\noindent
A reminder was sent on November the 6\textsuperscript{th} (three weeks after the first attempt) to all authors for which we did not receive a reply and for which the e-mail did not bounce. The communication contained the detailed query shown in~\ref{appendix:letter2}.\\

\noindent
Eight e-mails were sent a few weeks after the first iteration, to correct an error in the automated data acquisition. All responses received before December the 15\textsuperscript{th} $2017$ were considered in this survey.

\noindent
All e-mails were sent using the institutional e-mail address of one of the authors, as in other studies. The possibility for a biased reply-behavior when sending the e-mail as a student or as a professor was not addressed.

\section{Results}
\label{subsec:received_data}

\definecolor{lightgray}{gray}{0.9}

\begin{table*}[tb]
\centering
\resizebox{0.95\textwidth}{!}{
\rowcolors{2}{}{lightgray}
\begin{filecontents*}{data.csv}
Year,Bounce,Reply,No reply,Minutes,a,b,c,d,Check,Bouncep,Replyp,No replyp,Sum,Data
2000,7,3,1,1440,1,0,1,0,11,63.64,27.27,9.09,100,1
2001,7,0,4,0,0,0,0,0,11,63.64,0,36.36,100,0
2002,3,3,5,0,0,0,1,1,11,27.27,27.27,45.45,99.99,1
2003,3,0,8,0,0,0,1,1,11,27.27,0,72.73,100,0
2004,3,4,4,0,1,0,1,1,11,27.27,36.36,36.36,99.99,1
2005,1,1,9,120,0,0,0,0,11,9.09,9.09,81.82,100,0
2006,2,2,7,10,0,0,0,0,11,18.18,18.18,63.64,100,1
2007,0,2,9,0,1,0,1,1,11,0,18.18,81.82,100,0
2008,3,2,6,130,0,0,0,0,11,27.27,18.18,54.55,100,2
2009,2,2,7,0,1,0,0,0,11,18.18,18.18,63.64,100,0
2010,2,5,4,150,0,0,2,1,11,18.18,45.45,36.36,99.99,1
2011,5,2,4,10,0,0,0,0,11,45.45,18.18,36.36,99.99,0
2012,1,1,9,1,0,0,1,2,11,9.09,9.09,81.82,100,1
2013,2,4,5,60,0,0,0,0,11,18.18,36.36,45.45,99.99,0
2014,1,4,6,30,0,1,1,1,11,9.09,36.36,54.55,100,1
2015,0,0,11,0,0,0,0,0,11,0,0,100,100,0
2016,0,2,9,0,0,0,0,0,11,0,18.18,81.82,100,2
\textbf{Total},42,37,108,1951,4,1,9,8,187,22.46,19.79,57.75,100,11
\end{filecontents*}
\pgfplotstabletypeset[
col sep = comma,
columns={Year,Bounce,Bouncep,Reply,Replyp,No reply,No replyp,Minutes,a,b,c,d,Data},
every head row/.style={before row=\toprule%
\multicolumn{1}{c}{\textbf{Year}} & \multicolumn{2}{c}{\textbf{Bounces}} & \multicolumn{2}{c}{\textbf{Reply}} & \multicolumn{2}{c}{\textbf{No Reply}} & \multicolumn{1}{c}{\textbf{Minutes}} & \multicolumn{1}{c}{\textbf{Lost}} & \multicolumn{1}{c}{\textbf{Confidential}} & \multicolumn{1}{c}{\textbf{Reference}}& \multicolumn{1}{c}{\textbf{Others}} & \multicolumn{1}{c}{\textbf{Available}}\\
,after row=\midrule,
output empty row} ,
every last row/.style={after row=\bottomrule},
every nth row={17}{before row=\midrule},
display columns/0/.style={column name={}, string type, column type={l|}},
display columns/1/.style={column name={}, column type={c}},
display columns/2/.style={column name={}, column type={l|},postproc cell content/.style={@cell content=\textit{##1\%}}},
display columns/3/.style={column name={}, column type={c}},
display columns/4/.style={column name={}, column type={l|},postproc cell content/.style={@cell content=\textit{##1\%}}},
display columns/5/.style={column name={}, column type={c}},
display columns/6/.style={column name={}, column type={l|},postproc cell content/.style={@cell content=\textit{##1\%}}},
display columns/7/.style={column name={}, column type={l|}, string type},
display columns/8/.style={column name={}, column type={c}},
display columns/9/.style={column name={}, column type={c}},
display columns/10/.style={column name={}, column type={c}},
display columns/11/.style={column name={}, column type={c|}},
display columns/12/.style={column name={}, column type={c}},
columns/Data/.append style={string replace={0}{}},
columns/a/.append style={string replace={0}{}},
columns/b/.append style={string replace={0}{}},
columns/c/.append style={string replace={0}{}},
columns/d/.append style={string replace={0}{}},
columns/Minutes/.append style={string replace={0}{}},
]
{data.csv}
}
\caption{Analysis of the data obtained from the $187$ sent e-mails to the first author of the top-eleven cited papers from $2000$ to $2016$.}
\label{tab:data}
\end{table*}
Out of the $187$ papers selected, only one publication provided the experimental raw data attached as supplementary data on the journal's website.\\

\noindent
Table~\ref{tab:data} lists the data analysis for the $187$ e-mails sent. The first column presents the number of e-mails that bounced. The second column shows the number of replies to either the first or second e-mail. Note that there is no distinction between a positive or negative reply with respect to sharing the data. Only $30$ authors that did not respond to the first e-mail responded after receiving the second. The third column presents the number of no replies, which means that we did not obtain an error from the mail server and no answer $6$ weeks after sending the first e-mail.\\

\noindent
Table~\ref{tab:data} also lists the time, in minutes, required for the authors to retrieve the data (for those willing to share it). The following columns list the reasons the authors invoked for not providing the requested data. The last column lists the amount of available data sets. \\ 

\begin{figure*}[p]
     \centering
     \subfloat[][Number of non valid or non existing e-mail addresses per year for the $187$ corresponding authors contacted.]{\includegraphics[width=0.5\textwidth]{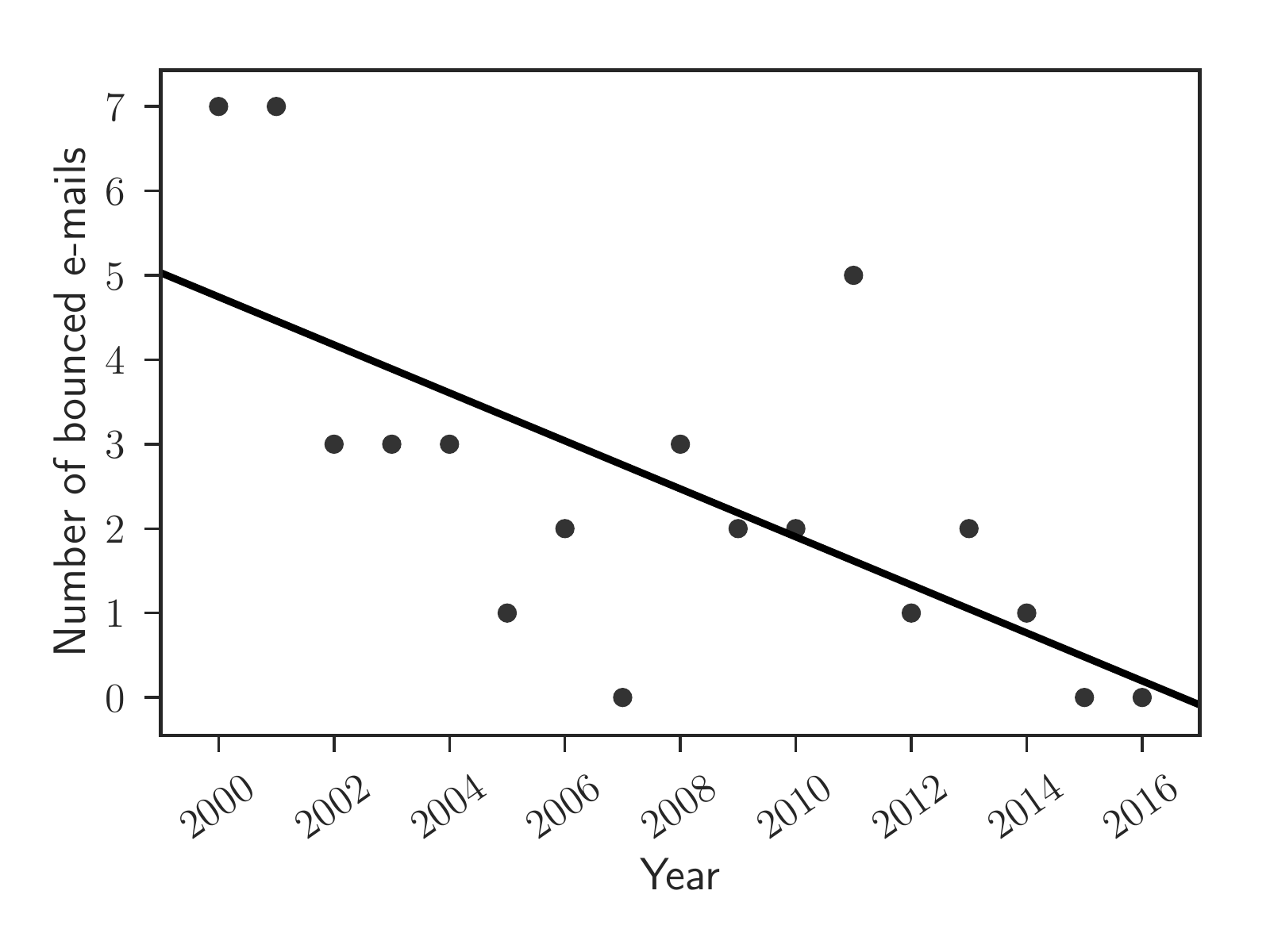}\label{fig:response:bounce}}~
     \subfloat[][Number of replies per year. This number includes the replies, where the authors reported that the data was lost or confidential.]{\includegraphics[width=0.5\textwidth]{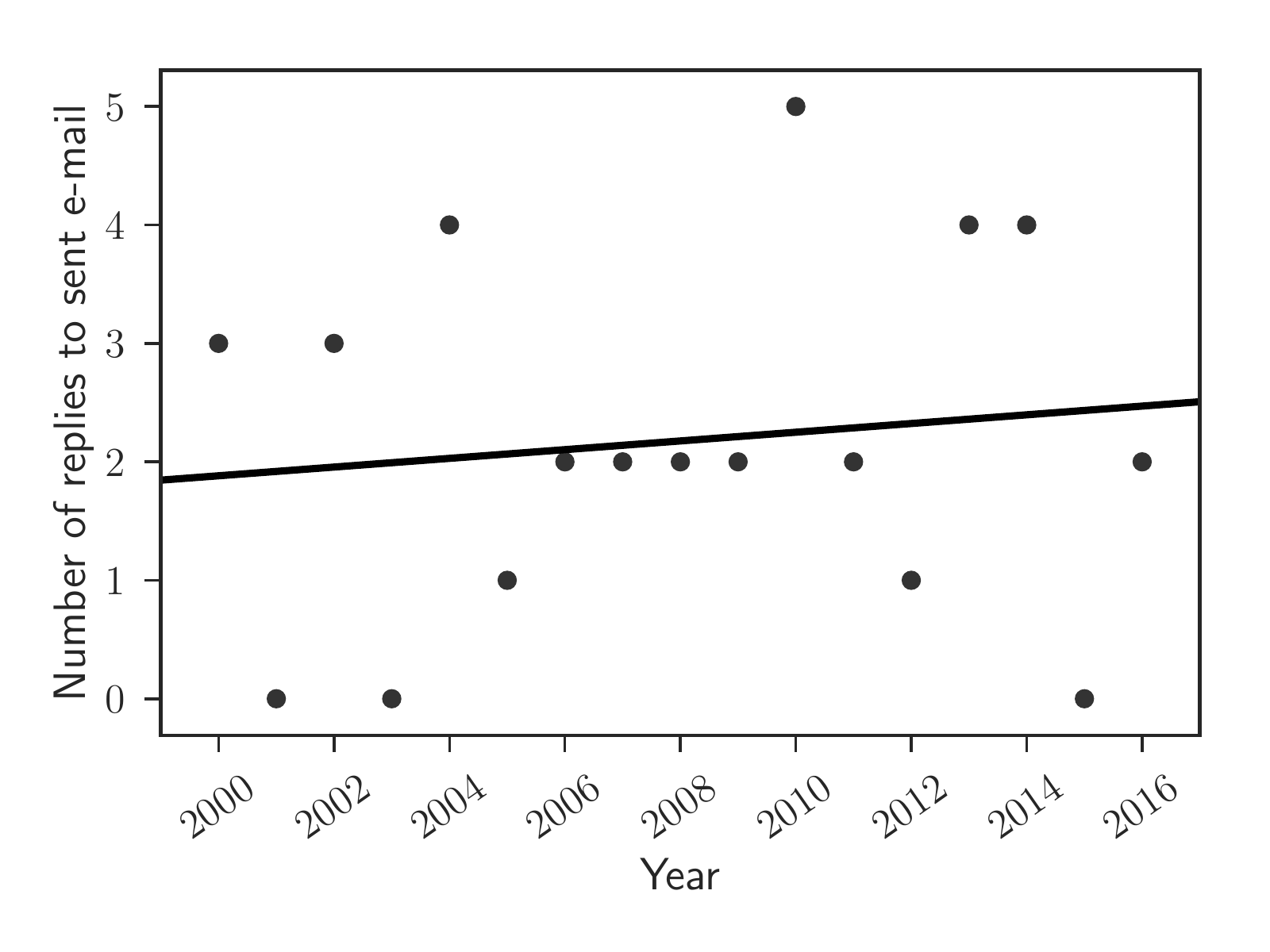}\label{fig:response:reply}}\\
     \subfloat[][Number of authors having working e-mail addresses who did not reply to the first or second e-mail.]{\includegraphics[width=0.5\textwidth]{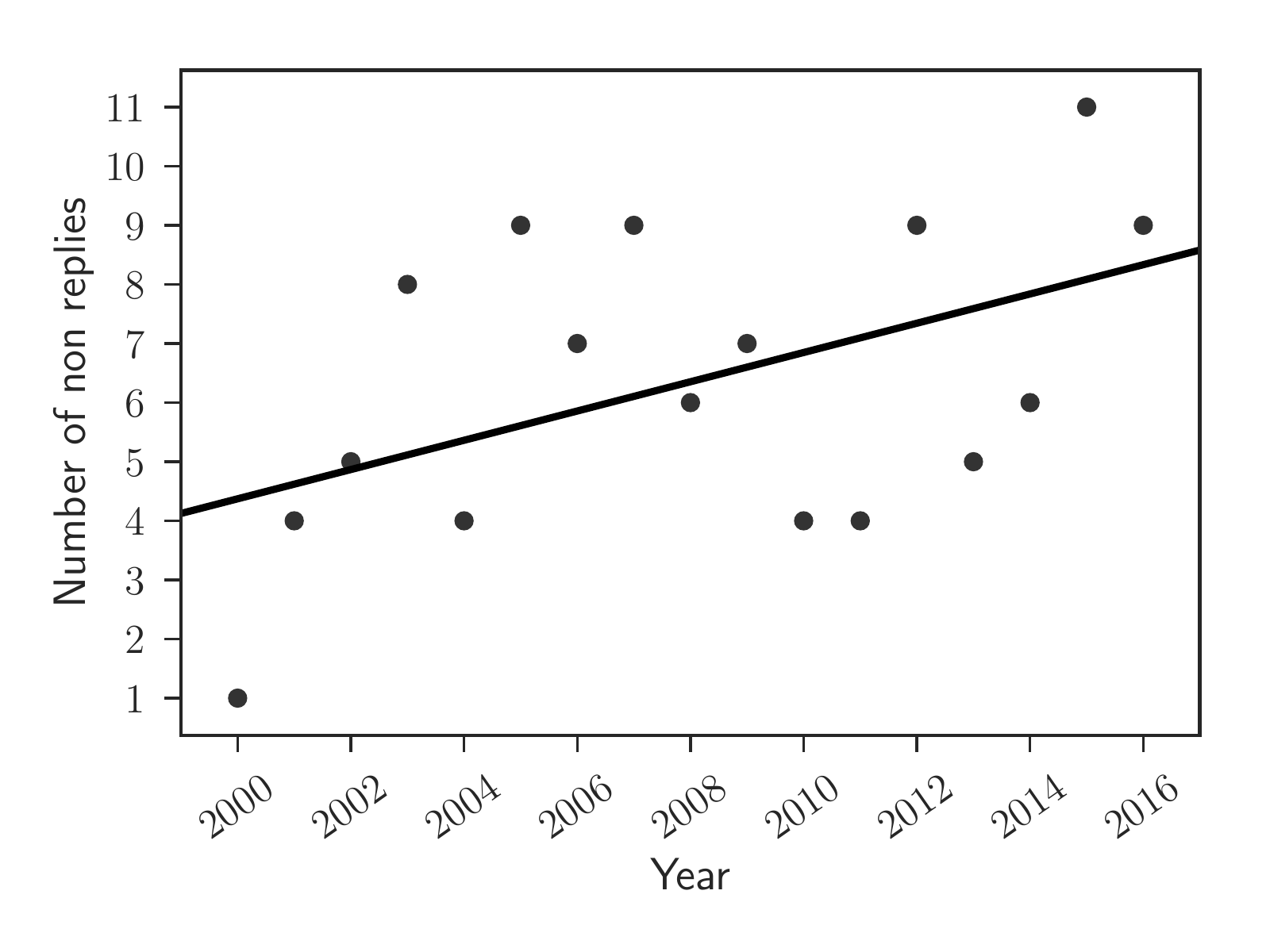}\label{fig:response:noreply}}~
     \subfloat[][Number of data sets the authors were willing to share. ]{\includegraphics[width=0.5\textwidth]{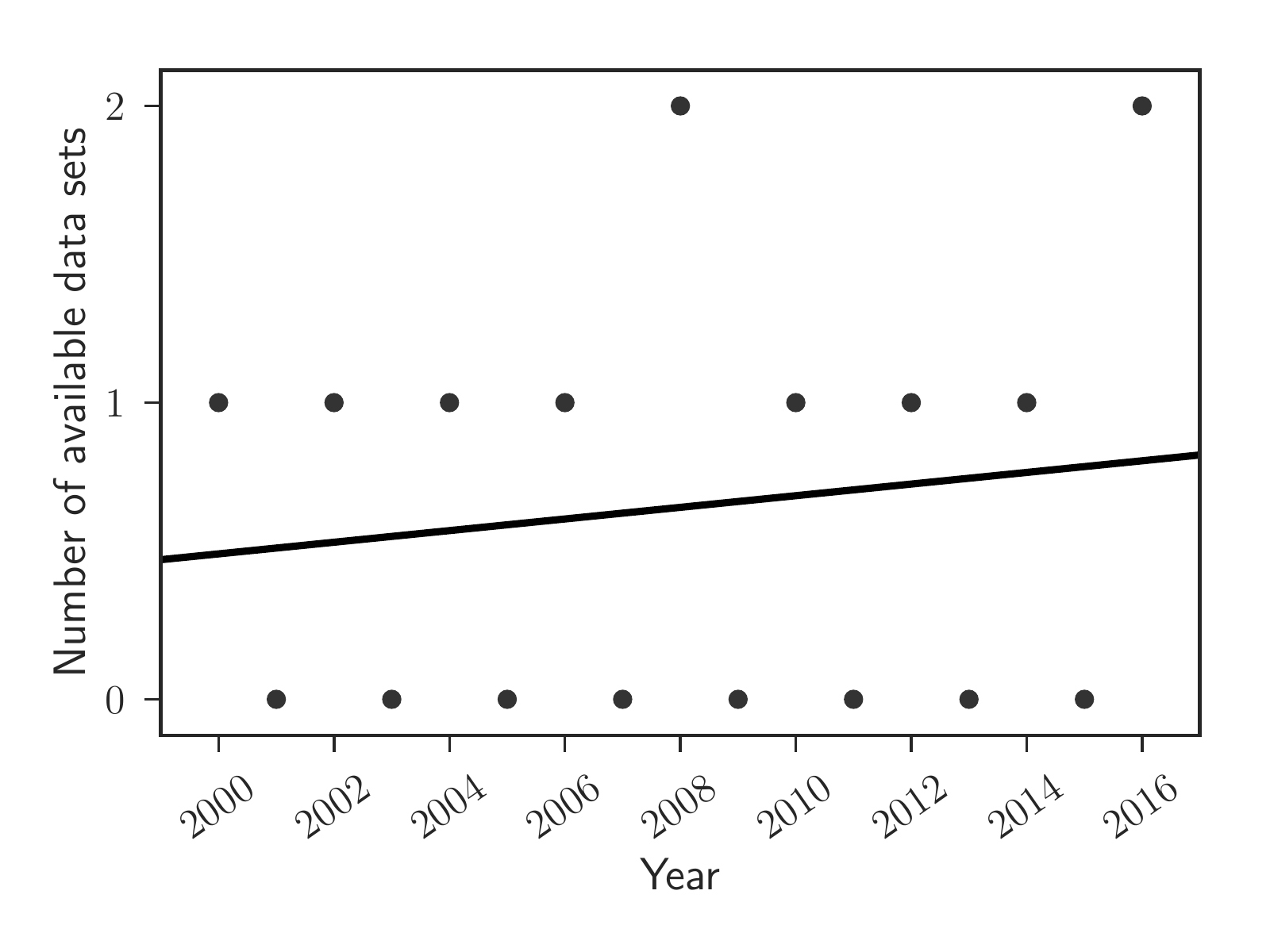}\label{fig:response:data}}
     \caption{Collected data with respect to the author responses as scatter plots. A linear regression (black line) for the collected data with respect to the responses of the authors was done. 
     }
     \label{fig:response:author}
\end{figure*}

\noindent
Figure~\ref{fig:response:author} shows the collected data with respect to the author responses as scatter plots. The linear regression between the year and the quantity of interest is presented. 
Figure~\ref{fig:response:bounce} shows the bounces for non valid or non existing e-mail addresses per year. Figure~\ref{fig:response:reply} presents the number of replies received to the first or second e-mail.
Figure~\ref{fig:response:noreply} presents the number of authors having working addresses who did not reply. Figure~\ref{fig:response:data} shows the number of times the requested data was available per year.

\section{Discussion}
\label{sec:discussion}

\subsection*{Outdated contact information}
Bouncing e-email addresses hindered the contact with the original authors and limited the acquisition of original research data. The number of bouncing e-mail addresses declines during the observation period, as seen in Figure~\ref{fig:response:bounce}. From $2000$ to $2002$, the average number of bouncing e-mails is $5.77$, while this number drops to $0.33$ in the $2014$--$2016$ period.

\subsection*{No reaction to our requests}
Figure~\ref{fig:response:noreply} shows that the number of authors who did not reply to our e-mails increased over the years while Figure~\ref{fig:response:bounce} shows that the number of invalid emails increased over the years. This observation suggests that authors who published more recently are less responsive than authors who published in previous years. 

\subsection*{Availability of raw experimental data}
Figure~\ref{fig:response:data} indicates that the availability is independent of the year and no trend within the linear regression could be found. The availability of data is generally low, with only $\data{Data}{17}$available data sets ($5.9$\%). 

\subsection*{Reasons invoked by the authors for not providing the data}
Authors were able to provide the reasons for not sharing the experimental data related to their publication. For example:
    \begin{enumerate}[noitemsep,nolistsep,label=(\alph*)]
        \item Retired author(s), or author(s) who left their institution, did not keep data backups;
        \item Author(s) have data storage plans and only keep large data sets for 5 to 10 years;
        \item Author(s) explained that sharing data would require work to render it usable by other researchers and they are not being paid to do so;
        \item One author explained that he believed that it is better for the experimentalist to do his own experiments.
    \end{enumerate}

\section{Conclusion and Outlook}
\label{sec:conclusion_and_outlook}
This work suggests that the availability of data sets in experimental fracture mechanics is very limited. Furthermore, the ability to retrieve the data is strongly attached to the possibility to contact the corresponding author. Retrieving the data becomes unlikely when the contact is lost with the corresponding author. Moreover, it seems that recent authors are less responsive to data sharing requests than authors who published in previous years.\\

\noindent
These facts limit the scientific capabilities of researchers to reproduce, build on, and check other scholars work. This study suggests that institutional e-mail addresses are insufficient means for obtaining experimental data sets. The lack of experimental data could also result from the fact that granting agencies and publishers do not require authors to make their raw data publicly available~\cite{spencer_thoughts_2010}. A possible solution to this issue could be the requirement to present a data management plan at the beginning of every new project. This can be required by research institutions or organizations that provide grants to researchers. It is also important to notice that providing the data is not sufficient: the data has also to be usable by other researchers. This means that the data must be labelled, explained, and put into context.\\

\noindent
We propose the following steps to improve experimental data availability:
\begin{itemize}[noitemsep]
	\item Requirement for ORCID~\cite{orcidURL}, 
    ResearcherID~\cite{researcheridURL}, 
    or other unique identifiers for publications that detach author(s) from their institutional e-mail addresses;
    \item Universities and other institutions listed as affiliations in scientific literature should provide forwarding e-mail addresses in case an author leaves their institution;
    \item Provide DOIs, like Zenodo~\cite{noauthor_zenodo:_2017} or Dataverse~\cite{noauthor_dataverse:_2017}, to make raw experimental data citable and provide more value for the academic curriculum. By making data sets citable, experimental researchers might take more time to prepare and store their data sets;
    \item Grant providing organizations should ensure the availability of experimental data by public funded projects, e.g.\ by asking for a data management plan.
\end{itemize}

As a future work, the authors of this publication would like to continue exploring the availability of experimental data through time by investigating the usability of data sets and determining guidelines to properly review an experimental data set in the field of mechanics, which could be done before updating a data set to one of the larger existing repositories.

\bibliographystyle{abbrv}
\bibliography{odem}

\appendix
\section*{Appendix}
\subsection{Data and Analysis}
The data and code for this study are available on github under following DOI: \url{10.5281/zenodo.1203766}.

\subsection{Initial email (sent 10/16/2017)}
\label{appendix:letter1}
Dear Prof. $<$author$>$\\\newline
My name is XXX and I am part of the Laboratory of Multi-Scale Mechanics at Polytechnique Montréal. We are currently working on a study aimed at determining how experimental data associated with publications changes through time.\\\newline 
We found your article $<$title$>$ among the 30 most cited articles on Scopus for the “experimental crack mechanics” query in $<$year$>$. We would be delighted if your publication could be part of our study. We are interested in the long term availability of raw experimental data and work supporting data, which was partly used in publications like yours. The complete study is anonymous and your response will not be used with your name or the reference in the study. For our study it would help if you could answer the following questions.
\begin{itemize}
\item Are you willing to share the experimental data with a peer to reproduce or to compare his simulations with the experiment?
\begin{itemize}
\item Could you also let us know how long (in minutes) it would take you to find the data? 
\end{itemize}
\item If your answer to the previous question is no, we would very much like to know the reason(s) behind that:
\begin{itemize}
\item The data is not available or lost 
\item The data is confidential 
\item Can you name a contact of the co-authors who can we ask for the data?
\item Other reasons (If you like please explain them)
\end{itemize}
\end{itemize}
If you have any further question on the design of this study or, are interested in its results, please feel free to contact us.\\\newline
Many thanks for your time and help. 

\subsection{Follow-up email (sent 11/06/2017 if no response to our initial email was received)}
\label{appendix:letter2}
Dear Prof. $<$author$>$\\\newline
I am following up on an e-mail we sent three weeks ago: we are a group of researchers from Polytechnique Montreal and the University of Stuttgart. Our researches are related to experimental mechanics or simulation and modeling in mechanics. We are currently trying to examine how the availability of experimental data in publications changes over time. Here, we are interested if the data is still available for reproducibility or the usage for benchmarks in simulation.\\\newline 
We handle your answer anonymous and your response will not be used with your name or the reference in the study. For our study it would really help if you could answer the questionnaire for your article $<$title$>$ published in $<$year$>$, it will take less than two minutes.  
\begin{itemize}
\item Are you willing to share the experimental data with a peer to reproduce or to compare his simulations with the experiment?
\begin{itemize}
\item Could you also let us know how long (in minutes) it would take you to find the data? 
\end{itemize}
\item If your answer to the previous question is no, we would very much like to know the reason(s) behind that:
\begin{itemize}
\item The data is not available or lost 
\item The data is confidential 
\item Can you name a contact of the co-authors who can we ask for the data?
\item Other reasons (If you like please explain them)
\end{itemize}
\end{itemize}
If you have any further question on the design of this study or, are interested in its results, please feel free to contact us or visit our project’s blog [0].\\\newline
Many thanks for your time and help. \\\newline
[0] https://opendataexpmechanics.github.io/
\end{document}